\documentclass[preprint,aps,pra,showpacs]{revtex4}
\usepackage{tabularx}
\usepackage{dcolumn}
\usepackage{graphics}
\usepackage{bm}
\usepackage[dvips]{graphicx}
\usepackage[dvips]{rotating}
\usepackage[latin1]{inputenc}
\usepackage{dcolumn}
\usepackage{tabularx}
\usepackage{amsmath}
\usepackage{amssymb}
\begin{document}
\draft

\title{Quantum over-barrier reflection effects manifested in the photodetachment cross sections }

\author{B. C. Yang and  M. L. Du}
\email{duml@itp.ac.cn} \affiliation{State Key Laboratory of
Theoretical Physics,Institute of Theoretical Physics, Chinese
Academy of Sciences, Beijing 100190, China}

\date{\today}

\begin{abstract}

Photodetachment of H$^-$ near a potential barrier is studied. A new
formula is presented for the cross sections induced by a barrier.
The new formula describes two quantum effects near barrier tops.
For energies near and above barrier tops, the quantum
over-barrier reflection effects are included and
the induced oscillations in the
cross sections are still prominent;
for energies near and below barrier tops, the quantum tunneling
across barriers is taken into account and consequently the oscillations are weakened. For energies far away from the barrier tops, the new formula agrees with the standard closed-orbit theory.
We demonstrate that a potential barrier of various width and location can be realized by placing a negative ion near a metal surface and applying an
electric field pointing to the surface. The width and location of the barrier can be systematically changed by varying the electric field strength and the distance between the negative ion and the surface. Results are also presented for estimating the sizes and locations of aforementioned quantum effects in the cross sections.

\par

\pacs{32.80.Gc, 03.65.Sq}

\end{abstract}

\maketitle

\section{introduction}

Photodetachment cross sections of negative ion in an external
electric field have been shown to exhibit oscillatory
structures\cite{Bryant,Gibson, Fabrikant1, Rau, Du1}. Similar oscillations occur in the presence of an interface\cite{Yang}. Studies on
the energy level shifts of Rydberg atoms and dynamics near a metal
surface have also been reported\cite{Ganesan, Simonovic, Wang, Hill,
Lloyd}. Inspired by these developments, the study of photodetachment
near a metal surface was proposed and the cross sections are shown
to be oscillatory because of the image charge\cite{Zhao}. Recently
we demonstrated that the photodetachment cross sections near the
metal surface can be systematically modified\cite{BCYang1} by
applying a weak electric field pointing away from the metal surface.
However, if the applied electric field points toward
the metal surface,  then potential barriers are created near the
negative ion. This new configuration implies interesting physics and
has not been studied for negative-ion photodetachment before.

Because quantum tunneling and quantum over-barrier reflection
effects are ignored, the standard closed-orbit theory\cite{Du2, Du3}
can not appropriately describe the photodetachment cross sections
for energies near the barrier tops. If the standard closed-orbit
theory is applied to our present system, the photodetachment cross
section would be oscillatory for energies below barrier tops; for
energies above the barrier tops, the oscillations would be absent.
Therefore, there is a sudden change in the cross sections at the
energies coinciding with the barrier tops. We note, in the atomic
case, it has been shown
that broader resonance states in the absorption spectra are associated
with quasi-classical orbits undergoing quantum above-barrier
reflection\cite{Delos1}.

In this paper, we study the photodetachment of a hydrogen negative
ion near a potential barrier created by a metal surface and an
electric field. We will derive a modified formula for the cross
sections, which requires the quantum reflection amplitude of the
barrier as an input. The modified formula gives an oscillatory cross
section for energy above barrier top induced by quantum over-barrier
reflection. It also gives a cross section with a weaker oscillation
for energy slightly below barrier top due to quantum tunneling. The
theoretical cross section is now smooth when the energy crosses the
barrier-top energy. As expected, for energies far away from the barrier tops, the modified formula agrees with the formula based on standard closed-orbit theory .

For potential barriers created by a metal surface and an electric
field, it is possible to vary the barrier width and location by
varying the electric field strength $F$ and the distance $d$ between the
negative ion and the metal surface. We will survey the parameter
space $(F,d)$ and present a guide which can be used to estimate the location
and the size of the energy window in which the quantum reflection
and tunneling effects are important. We find that the size of the
window is determined by the electric field strength alone and is
given accurately by $4F^{3/4}$. Atomic units are used unless
specified otherwise.

\section{hamiltonian and potential barriers}

The configuration for creating a potential barrier is depicted
schematically in Fig.1 where a negative ion H$^-$ sits in front of a
metal surface and an electric field $F$ is applied pointing to the
surface. We assume the negative ion sits at the coordinate origin (z=0). The distance between the negative ion and the metal surface is denoted as $d$.
The metal surface cuts the z-axis at z=-d.

Let $p_\rho$ and $p_z$ represent the momentum components of the electron in $\rho$ and $z$ directions in cylindrical coordinate system, then the
Hamiltonian governing the motion of the detached electron can be
written using the image method\cite{Ganesan,Zhao} as
\begin{equation}\label{}
     H=\frac{1}{2}(p_\rho^2+p_z^2)-\frac{1}{4(d+z)}+\frac{1}{4d}-Fz,
\end{equation}
where the second term represents the image potential induced by the
metal surface and the last term is the potential of the applied
electric field. We have also added a constant term $\frac{1}{4d}$ so
that the potential is set to zero at the origin. For the potential
in Eq.(1) and a weak electric field satisfying $F<\frac{1}{4d^2}$, a
fatter barrier is created on the positive z side;
when a strong electric field satisfying $F>\frac{1}{4d^2}$ is
applied, a thinner barrier is created lying between the surface and
the negative ion. Because the quantum over-barrier reflection
effects and quantum tunneling effects are stronger for the thinner
potential barriers, we will focus on the thinner barriers
($F>\frac{1}{4d^2}$) in the following discussions. It is easy to
show that the peak position of the thinner potential barrier is
located at $z_{max}=\sqrt{1/4F}-d$ and the barrier top energy is
$V_{max}=-\sqrt{F}+Fd+\frac{1}{4d}$. Furthermore, we find that the
width of the barrier is determined by the electric field alone as
will be discussed later. Thus the width of the barrier can be varied
by varying the electric field strength $F$, and the location of the
barrier in this system can be controlled by $d$ and $F$. A typical
thin potential barrier is shown in Fig.2.

The physical picture for the photodetachment of H$^-$near the
barrier top can be described as the following. When a photon is
absorbed by the negative ion, the loosely bound $s$-state electron
goes into an outgoing $p$-wave, propagating out from the negative
ion in all directions. When the outgoing wave propagating initially
toward the metal surface reaches the potential barrier, part of the
wave is reflected by the potential barrier and the reflected wave
then returns to the region of the negative ion. The interference
between the returning wave and the outgoing wave induces
oscillations in the cross sections.

If we assume the metal surface absorbs the detached-electron reaching the surface and the standard closed-orbit theory is applied for the present
system\cite{Zhao}, the photodetachment cross section for photon polarized in the z direction would be given by the following formula,
\begin{equation}\label{}
    \sigma_{standard}=\frac{16\pi^2\sqrt{2}B^2E^{3/2}}{3c(E_b+E)^3}+
    \Theta(V_{max}-E)\frac{8\pi^2B^2\sqrt{2E}}
    {c(E_b+E)^3T_{c}}\cos[S_{c}],~E\geq0,
\end{equation}
where\cite{Du1} $B=0.31552$ is related to the normalization of the
initial bound state of hydrogen negative ion; $c$ is the speed of
light in a.u. and is approximately 137; $E_b$ is the binding energy
of H$^-$ and is approximately $0.7542$eV; $E$ is the initial kinetic
energy of the detached electron. The sum of $E$ and $E_b$ is equal
to the photon energy. $T_{c}$ and $S_{c}$ are, respectively, the
classical transit time and the action for the electron along the
closed-orbit which goes along the z-axis from the origin to the
barrier and back to the origin for energy below barrier top.
$\Theta(x)$ is the Heaviside step function,
\begin{equation}\label{}
    \Theta(x)=\left\{\begin{array}{c}
              ~0,~~~~x<0; \\
              ~1,~~~~x\geq0.
            \end{array}\right.
\end{equation}

However, the standard treatment leading to the formula in Eq.(2)
ignores the quantum over-barrier reflection effects and quantum
tunneling effects and therefore is incorrect for energy close to the
barrier top. According to Eq.(2), when the energy is above the
barrier top energy $V_{max}$, the wave will not be reflected al all
and consequently there is no oscillation in the photodetachment
cross section. And when the energy is below the barrier top, the
wave is completely reflected in the standard treatment, giving an
overestimated oscillation. The incorrect description of the reflected wave
above and below the barrier top leads to a sudden jump in the
photodetachment cross section at the  barrier top energy. Our
purpose in this article is to derive a formula which includes both
the quantum over-barrier reflection effects and the quantum
tunneling effects and to explicitly demonstrate the effects for the
system described in Eq.(1).

\section{derivation of the modified formula}

We start with a quite general expression for the photodetachment
cross section\cite{Du1, Du-para},
\begin{equation}\label{}
\sigma=\frac{4\pi^2}{c(E+E_b)}\sum_n\int|\langle
    f_{E,n}\mid p_z \mid i\rangle|^2\delta(E(f)-E)df,
\end{equation}
where the laser polarization is assumed along z-axis, $|i\rangle$
denotes the initial wave function and $|f_{E,n}\rangle$ represents
the final wave function of the n-th  channel. The summation includes
all the involved final channels.

The initial bound state wave function of H$^-$ is usually described
as $\psi_i=B\frac{e^{-k_br}}{r}$\cite{Du1}, where $k_b=\sqrt{2E_b}$,
both $B$ and $E_b$ are known constants. For this initial wave
function and noting the potential in Eq.(1) depends on z but not x and
y, the general expression in Eq.(4) can be turned into the
following form with some manipulations given in the Appendix A
\begin{equation}\label{}
    \sigma=\frac{8\pi^3B^2}{c(E+E_b)^3}
    \sum_{n=1}^2\int^E_{-\infty}\big|\{\frac{\partial\langle z|E_z, n\rangle}{\partial
    z}\}_{z=0}\big|^2dE_z,
\end{equation}
where $\langle z|E_z, n\rangle$ is the energy-normalized
coordinate-space wave function in the z direction.

For the present problem with a potential barrier, there are two
channels. The $n=1$ channel  represents a wave with incoming and
reflected waves on the left side of the barrier and transmitted wave
on the right side of the barrier; the $n=2$ channel represents a
wave with incoming and reflected waves on the right side of the
barrier and a transmitted wave on the left side of the barrier.

The wave function for channel $n=1$ far away from the barrier and outside the surface can be written as
\begin{eqnarray}
  \nonumber \langle z|E_z, n=1\rangle &=& \frac{1}{\sqrt{2\pi k_z}}\exp(i\int_0^zk_zdz')+\frac{\alpha}{\sqrt{2\pi k_z}}\exp(-i\int_0^zk_zdz'), -d<z\ll z_{max};\\
  & & \frac{\eta}{\sqrt{2\pi k_z}}\exp(i\int_0^zk_zdz'),z \gg z_{max}.
\end{eqnarray}
where $k_z=\sqrt{2(E_z-V(z))}$; $\alpha$ and $\eta$ are respectively
the quantum reflection amplitude and the quantum transmission
amplitude of the barrier with an incoming wave from the left.

For channel $n=2$, the wave far away from the barrier and outside the surface can be written as
\begin{eqnarray}
  \langle z|E_z, n=2\rangle &=& \frac{1}{\sqrt{2\pi k_z}}\exp(-i\int_0^zk_zdz')+\frac{\gamma}{\sqrt{2\pi k_z}}\exp(i\int_0^zk_zdz'), z \gg z_{max};\\
  \nonumber  & & \frac{\beta}{\sqrt{2\pi k_z}}\exp(-i\int_0^zk_zdz'),
  -d<z \ll z_{max}.
\end{eqnarray}
where $\gamma$ and $\beta$ are respectively the quantum reflection
amplitude and the quantum transmission amplitude of the barrier with
an incoming wave from the right. At the same scattering energy,
$\eta$ and $\gamma$ satisfy the following relationship\cite{Landau},
\begin{equation}\label{}
    |\eta|^2+|\gamma|^2=1.
\end{equation}

Eq.(5) can now be evaluated using the expressions in Eq.(6) and Eq.(7) to get the contributions corresponding to the two channels. They are
\begin{eqnarray}
  \sigma_1 &=& \frac{4\pi^2B^2}{c(E+E_b)^3}\int^E_0k_0|\eta|^{2}dE_z, \\
  \sigma_2 &=&
  \frac{4\pi^2B^2}{c(E+E_b)^3}\int^E_0k_0
  (1+|\gamma|^{2})dE_z-\frac{8\pi^2B^2}
  {c(E+E_b)^3}\int^E_0\sqrt{2E_z}Re(\gamma)dE_z,
\end{eqnarray}
where $k_0=\sqrt{2(E_z-V(z=0))}$ and the lower limit of the
integration is taken to be zero to be consistent with the
semiclassical approximation of the wave function near the negative
ion. The total photodetachment cross section is obtained by summing
the contributions in Eq.(9) and Eq.(10),
\begin{equation}\label{}
    \sigma=\frac{16\sqrt{2}\pi^2B^2E^{3/2}}{3c(E+E_b)^3}-\frac{8\pi^2B^2}{c(E+E_b)^3}\int^E_0\sqrt{2E_z}Re(\gamma)dE_z.
\end{equation}
Now we define $\gamma=|\gamma|e^{i\phi}$, then we can write
$\int^E_0\sqrt{2E_z}Re(\gamma)dE_z=Re\int^E_0\sqrt{2E_z}
    |\gamma|e^{i\phi}dE_z.$

For the present system, we find $|\gamma(E)|$ and $\phi(E)$ ($\phi(E)$ is related to $S_q(E)$ by Eq.(14),see below) behave like those displayed in Fig.3. The phase $\phi(E)$ increases with energy and there is no stationary point in the interval (0,E). Assuming $\frac{d\phi(E)}{dE}$ is large,we argue the main contribution of the above integration is from the upper boundary,
\begin{equation}\label{}
Re\int^E_0\sqrt{2E_z}|\gamma|e^{i\phi}dE_z\approx
    Re\frac{\sqrt{2E}|\gamma(E)|e^{i\phi(E)}}{iT_q(E)},
\end{equation}
where $T_q(E)=\frac{d\phi(E)}{dE}$. Based on an analysis guided by closed-orbit theory\cite{Du2, Du3}, it is concluded that the contribution to the integral in Eq.(12) from the lower boundary should be neglected and when the approximation in Eq.(12) is used in Eq.(11), we obtain the modified photodetachment cross section near a barrier,
\begin{equation}\label{}
    \sigma=\frac{16\sqrt{2}\pi^2B^2E^{3/2}}{3c(E_b+E)^3}+
    |\gamma(E)|\frac{8\pi^2B^2\sqrt{2E}}
    {c(E_b+E)^3T_q(E)}\cos[S_q(E)],
\end{equation}
where,
\begin{eqnarray}
  S_q(E) &=& \phi(E)+\pi/2; \\
  T_q(E) &=& \frac{d\phi(E)}{d E}.
\end{eqnarray}
By comparing with the numerical calculations using Eq.(5), we have verified that the approximate formula in Eq.(13) is accurate.

We note that for
energy well below the barrier top,$|\gamma|$ approaches unit,
Eq.(13) then should go back to the standard formula Eq.(2). This
implies, in the low energy limit, $S_q$ should be equal to
$S_c=\oint pdq$, and $T_q$ should be equal to the classical transit
time $T_c$. However, for energy above the barrier top, both $T_c$
and $S_c$ are not defined, while $S_q$,$T_q$ and $|\gamma|$ are all
well defined by quantum scattering theory. For energy below the
barrier top, because $|\gamma|$ is smaller than one due to quantum
tunneling across the potential barrier, the oscillation amplitude
described by Eq.(13) is smaller compared to that in Eq.(2). For
energy above the barrier top, the modified formula in Eq.(13) still
describes an oscillation in the cross section associated with the
quantum over-barrier reflection effects. The cross section described
by Eq.(13) is smooth when the detached-electron energy crosses the
barrier top energy $V_{max}$.

\section{numerical calculations}

We now apply Eq.(13) to the Hamiltonian in Eq.(1). For $d=200a_0$
and $F=300$kV/cm, the potential barrier is illustrated in Fig.2. An
iterative numerical method was applied for solving the quantum
scattering problem of the barrier\cite{Du4} with the boundary conditions
given in Eq.(6) and Eq.(7) for $z>-d$, assuming the detached-electron is absorbed when it reaches the surface. The calculated
$|\gamma|$ and $S_q(E)$ are shown in Fig.3 as functions of photon
energy. The important energy window we are focusing on near the
barrier top is marked gray in both Fig.2 and Fig.3. We observe, as
the energy is increased crossing the energy window from below to
above the barrier top, the amplitude $\gamma$ displayed in Fig.3(a)
decreases and the phase $S_q(E)$ shown in Fig.3(b) increases
smoothly. Fig.3(b) also demonstrates that the phase $S_c(E)$ (dotted
lines) and $S_q(E)$ coincide for energy sufficiently below the
barrier top.

The photodetachment cross sections with the potential barrier
displayed in Fig.2 are calculated and shown as the red solid lines
in Fig.4. For the purpose of comparison, we also show the results
calculated using the standard formula in Eq.(2). The light dotted
lines represent the cross sections of a free negative ion without
the metal surface and the electric field. The rectangular area in
Fig.4(a) corresponds to the gray energy region near the barrier top
in Fig.2 and Fig.3. For clarity, it is amplified and shown in
Fig.4(b). In these figures, we observe the cross sections described
by Eq.(2) has a sudden jump at the barrier top because it neglects
the quantum over barrier reflection effects and quantum tunneling
effects. In contrast, the cross sections described by the modified
formula in Eq.(13) are smooth and well-behaved. The oscillations in
the cross sections above the barrier top are now clearly visible and
vary smoothly as the energy decreases across the barrier top. The
oscillations in the cross sections above barrier top are the direct
manifestations of the quantum over-barrier reflection effects.

Similar calculations can be carried out for other barriers.
For example, in Fig.5 we show the cross sections for
$d=400a_0$ and $F=300$kV/cm. This case corresponds to an energy
window located at higher energy. There are more visible oscillations in the
region above the barrier top in the cross sections.

\section{sketches of parameter space}

For the Hamiltonian system generated by a metal surface and an
electric field in Eq.(1), a potential barrier is always created
between the metal surface and the negative ion for any pair of
$(d,F)$ satisfying  $F>\frac{1}{4d^2}$. The location and the width
of the barrier of course depend on the values of $d$ and $F$ and
they in turn control the location and size of the energy window in
the cross sections related to the quantum over-barrier reflection
effects and quantum tunneling effects. In this section we present
some formulas and figures so one can quickly  estimate the barrier
scattering region in the cross sections for any given $d$ and $F$.

First, as we mentioned, the barrier top corresponds to an
detached-electron energy $V_{max}=-\sqrt{F}+Fd+\frac{1}{4d}$.
Therefore, the position of the barrier scattering region (such as
the shaded energy range in Fig.2 or the rectangular region in
Fig.4(a)) in the cross section is located at photon energy
$E_{c}=E_b+V_{max}$. In Fig.6(a), we show how $E_{c}$ depends on $d$
and $F$. In the figure, each solid line corresponds to the same
value of $E_{c}$ given on the right in eV. The locations of the two
calculations in the previous section are illustrated as circles.

Next, we estimate the energy width of the barrier scattering region or the rectangular region in Fig.4 or Fig.5.
As illustrated in Fig.3, the energy width corresponds to the
reflection coefficient varying from almost one to almost zero or the
transmission coefficient varying from almost zero to almost one. To
proceed, let us approximate the potential barrier near the barrier
top by $-K(z-z_{max})^2/2+V_{max}$. It is easy to show that
$K=4F^{3/2}$. The transmission coefficient for this parabolic
potential can be written as\cite{Kemble}
\begin{equation}
    \mathbb{T}=\left \{
   \begin{array}{cc}
    \frac{1}{1+\exp(-2\pi|E-V_{max}|/\sqrt{K})}, &{\rm if}~~E\geq V_{max};\\
    \frac{1}{1+\exp(2\pi|E-V_{max}|/\sqrt{K})},
    &{\rm if}~~E \leq V_{max}.
    \end{array}
    \right.
\end{equation}
The energy width $\Delta E$ can be simply estimated as corresponding
to the transmission coefficient in the range between
$\frac{1}{1+e^{2\pi}}$ ($\sim0.0019)$ and $\frac{1}{1+e^{-2\pi}}$
($\sim0.9981$). Then we have the result
\begin{equation}\label{}
    \Delta E = 2\sqrt{K} = 4F^{\frac{3}{4}}.
\end{equation}
Fig.6(b) shows the energy width $\Delta E$ as a function of $F$.

The fact that the width $\Delta E$ depends only on the electric
field can be understood in the following way. In fact, all the
barriers corresponding to the same $F$ but different $d$ have
\emph{the same shape}. To see this, we use the new coordinate
$z_s=z+d$ relative to the metal surface in the potential
$V(z,d,F)=-\frac{1}{4(d+z)}+\frac{1}{4d}-Fz$ in Eq.(1). We then have
$V(z,d,F)=(-1/(4z_s)-Fz_s)+(Fd+\frac{1}{4d})$. The shape of the
potential $V(z,d,F)$ is determined by $(-1/(4z_s)-Fz_s)$ which is
independent of $d$.

Finally, we estimate the number of oscillations in the barrier
scattering window given in Eq.(17). The number of full oscillations
in the energy window can be obtained from the phase change $\Delta
\phi$ across the energy window divided by $2\pi$. The phase change
$\Delta \phi$ is approximately twice the semiclassical phase change
from the barrier top to the lower edge of the energy window, $\Delta
\phi=2[S_c(V_{max})-S_c(V_{max}-\Delta E/2)]$, where $S_c(E)=\oint
p(z)dz$ is the integral along the closed-orbit which goes out from
the negative ion to the barrier and back to the negative ion. For
the potential in Eq.(1), we have
\begin{equation}\label{}
    S_c=\frac{4\sqrt{2F}}{3}\bigg\{\sqrt{z_m+d}\big[(z_m+z_n+2d)\mathcal{E}(\mu,\lambda)
    -(z_m-z_n)\mathcal{F}(\mu,\lambda)\big]-(z_m+2z_n+2d)\sqrt{\frac{z_m}{z_n}d}\bigg\}~,
\end{equation}
where $z_m$ and $z_n$ are the two turning points of the barrier. They are given by
\begin{eqnarray}
  z_m &=& \frac{1}{2F}\bigg[-(Fd+E-\frac{1}{4d})+\sqrt{\big(Fd-(E-\frac{1}{4d})\big)^2-F}\bigg]; \\
  z_n &=& \frac{1}{2F}\bigg[-(Fd+E-\frac{1}{4d})-\sqrt{\big(Fd-(E-\frac{1}{4d})\big)^2-F}\bigg].
\end{eqnarray}
$\mathcal{F}(\mu,\lambda)$ and $\mathcal{E}(\mu,\lambda)$ are, respectively,
the elliptic integrals of the first kind and the second kind\cite{table}. The two parameters are defined as
\begin{eqnarray}\label{}
    \mu &=& \arcsin\sqrt{\frac{z_m}{z_n}}~,0<\mu\leq\frac{\pi}{2};\nonumber\\
    \lambda &=& \sqrt{\frac{z_n+d}{z_m+d}}~,0<\lambda\leq 1.
\end{eqnarray}

In Fig.6(c) we show the number of full oscillations as function of
$d$ and $F$. The estimated full oscillations for the two
calculations in Sec.IV are respectively $2.05$ and $3.30$, which are
consistent with the cross sections displayed in Fig.4 and Fig.5.
Fig.6(a),Fig.6(b) and Fig.6(c) together provide a quick and useful
estimate of the location, the size and the number of oscillations
for the barrier scattering energy window in the cross sections for
any given $d$ and $F$.

\section{conclusions}

We studied photodetachment cross sections for a negative ion near
potential barriers. The potential barriers can be created with a metal surface and  an electric field pointing to it. The image potential should be accurate when the distance from the metal surface is greater than about ten atomic units and if the kinetic energy of detached-electron is less than the Fermi energy of the metal\cite{Jones}.
We have extended closed-orbit theory
and derived a modified formula in Eq.(13). This new formula describes both
quantum over-barrier reflection effects and quantum tunneling
effects. These effects have been excluded in the standard formula in
Eq.(2). As illustrations, we have calculated cross sections for two different barriers and shown them in Fig.4 and Fig.5. The most significant features in the
cross sections are the visible oscillations above the barrier tops. Such
oscillations are direct manifestations of the quantum over-barrier
reflection effect in the photodetachment cross sections and may be
observable in experiments. One may imagine an experimental setup in which a beam of negative ions travels quickly through the interaction region from left to right as in Fig.1, keeping the distances between the ions and the metal surface close to a certain value. While the ions are in the interaction region, a laser is applied
and the resulting detached-electron is measured.

For the combined electric field and metal surface system in Eq.(1),
we have surveyed the parameter space $(d,F)$ satisfying
$F>\frac{1}{4d^2}$. We have estimated the location and the width of
the barrier scattering region in the cross sections. The
oscillations inside the barrier scattering region have been
estimated as well. The results are comprehensively summarized in
Fig.6. For any pair of $d$ and $F$ one can use the map given in
Fig.6 to find the barrier scattering window without doing
complicated calculations. For future experimental reference, we note
the electric field $F$ alone determines the potential barrier shape
and width.

\begin{center}
{\bf ACKNOWLEDGMENT}
\end{center}
\vskip8pt This work was supported by NSFC grant No. 11074260.

\appendix

\section{}
In momentum space the initial-state wave function is
\cite{Du1}
\begin{equation}\label{1}
    \langle \mathbf{p}|i\rangle=2^{1/2}\pi^{-1/2}\cdot\frac{B}{k^2_b+p^2}
\end{equation}
and the final-state wave functions are
\begin{equation}\label{2}
    \langle \mathbf{p}|f_{E',n}\rangle=\delta(p_x-p'_x)\delta(p_y-p'_y)\langle p_z|E'_z,n\rangle.
\end{equation}
The matrix element in Eq.(4) can therefore be written as
\begin{equation}\label{3}
    \langle i|p_z|f_{E', n}\rangle=\frac{\sqrt{2}B}{\sqrt{\pi}}\int\frac{p_z\langle
    p_z|E'_z, n\rangle}{k^2_b+p'^2_x+p'^2_y+p^2_z}dp_z.
\end{equation}
Recognizing the main contribution to the cross section comes from the energy shell $E$, the slowly varying denominator is evaluated at the final energy surface and then moved out of the integral as was done earlier\cite{Du1}. The integral can then be simplified to
\begin{equation}\label{4}
    \langle i|p_z|f_{E', n}\rangle=\frac{\sqrt{2}B\pi^{-1/2}}{k^2_b+p'^2}\int p_z\langle
    p_z|E'_z, n\rangle dp_z.
\end{equation}
Because
\begin{equation}\label{5}
    \langle p_z|E'_z, n\rangle=\int\langle p_z|z \rangle \langle z|E'_z, n\rangle dz
\end{equation}
and
\begin{equation}\label{6}
    \langle p_z|z\rangle=\frac{1}{\sqrt{2\pi}}\exp(-ip_zz),
\end{equation}
the integral in Eq.(A5) can be written as
\begin{equation}\label{7}
    \int p_z\langle p_z|E'_z, n\rangle dp_z=i\cdot\sqrt{2\pi}\int
    \frac{d\delta(z)}{dz}\cdot\langle z|E'_z, n\rangle dz.
\end{equation}
Thus we have
\begin{equation}\label{8}
    \langle i|p_z|f_{E', n}\rangle=-i\cdot\frac{2B}{k^2_b+p'^2}\{\frac{\partial\langle z|E'_z, n\rangle}{\partial
    z}\}_{z=0}.
\end{equation}
When the result in Eq.(A8) is used in Eq.(4), we obtain Eq.(5).


\newpage


\begin{figure}[H]
  \includegraphics[width=250pt]{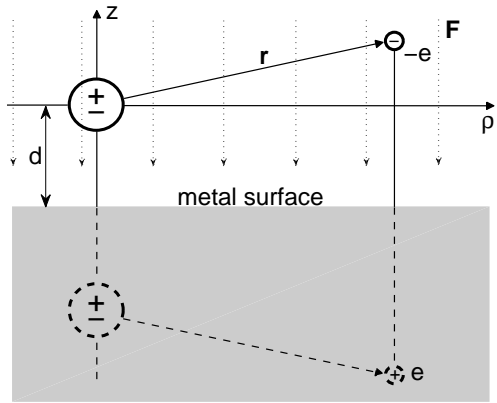}
  \caption{Schematic representation of the photodetachment process of H$^-$ in the presence of a metal surface and a static
electric field pointing to the surface. The real large circle represents the Hydrogen atom and the real small circle represents the detached-electron. The dotted-circles are the correspond images of the Hydrogen atom and the detached-electron in the metal. The distance between the negative ion and the metal surface is $d$. A potential barrier for the detached-electron is created between the negative ion and the metal surface for $F>\frac{1}{4d^2}$.  }
\end{figure}

\begin{figure}[H]
  \includegraphics[width=250pt]{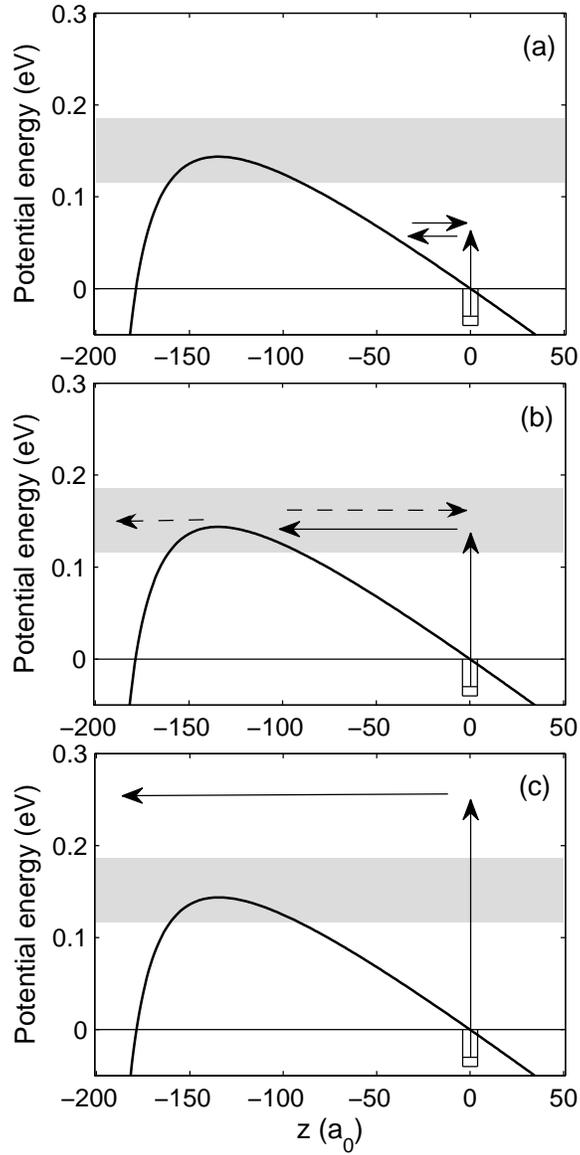}
  \caption{Illustration of the reflected waves in the photodetachment near a potential barrier created with an electric field $F=300$kV/cm and $d=200a_0$ ($a_0$ is the Bohr radius) in Eq.(1). (a) When the detached-electron energy is much lower than the barrier top, the barrier reflection is nearly complete;
  (b) in the gray region near the barrier top, quantum tunneling effects and quantum over-barrier reflection effects are significant,
  resulting in a partial reflection of the wave;
  (c) when the detached-electron energy is much higher than the barrier top, the reflection is negligible. }
\end{figure}

\begin{figure}[H]
  \includegraphics[width=250pt]{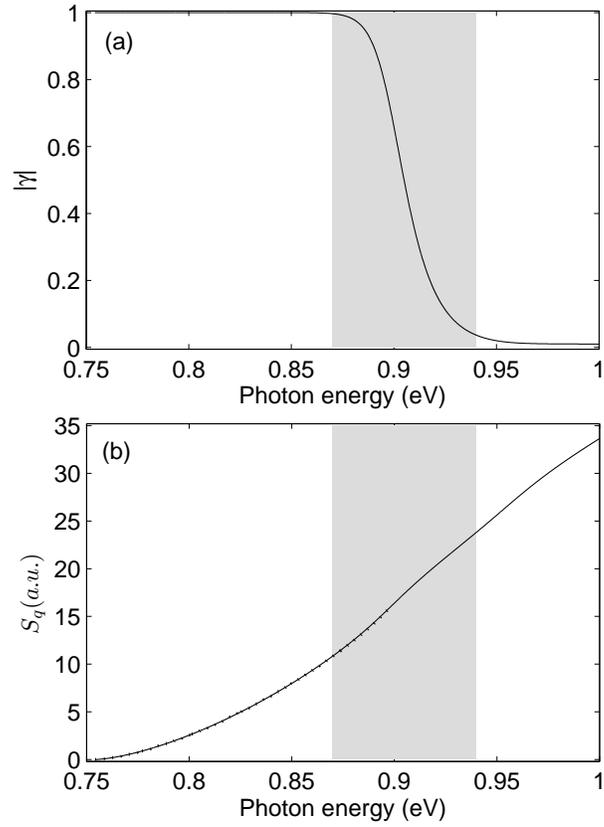}
  \caption{(a) Calculated $|\gamma|$ for the potential barrier in Fig.2;
  (b) calculated quantum phase $S_q(E)$ (solid line) is compared with
  the classical action $S_c(E)$ (dotted line).
  The gray region is the same as that in Fig.2.}
\end{figure}

\begin{figure}[H]
  \includegraphics[width=250pt]{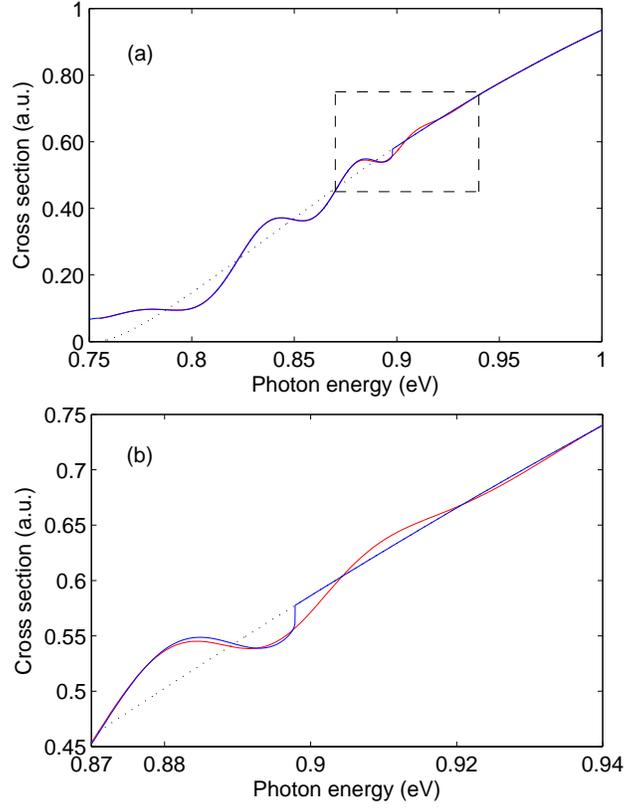}
  \caption{(Color online) Photodetachment cross sections (red curves) using the formula in Eq.(13) for the barrier in Fig.2. For comparison,
  the results obtained by Eq.(2) are also shown as blue curves.  The dotted lines are the cross sections of a free negative ion.
  The rectangular area in (a) corresponds to the gray energy window in Fig.2, which is amplified in (b) for clarity. }
\end{figure}

\begin{figure}[H]
  \includegraphics[width=250pt]{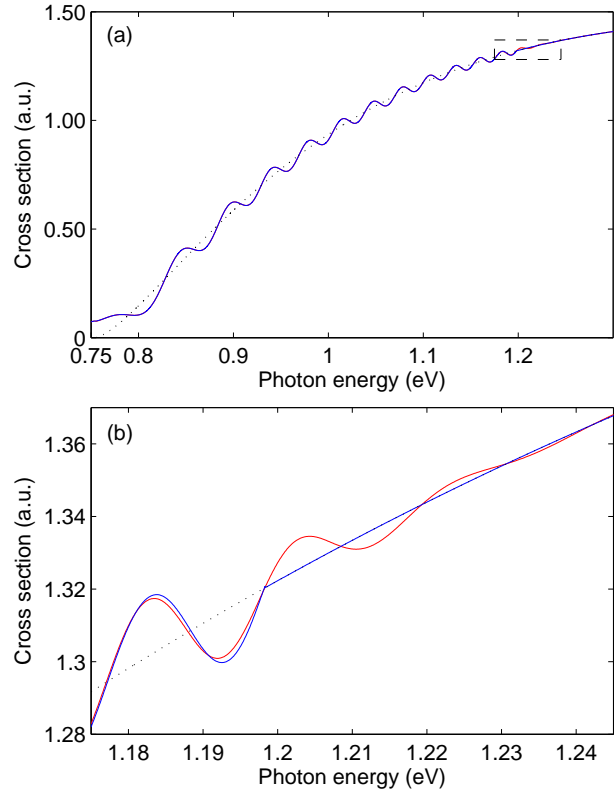}
  \caption{(Color online) Similar to Fig.4 but now $d=400a_0$ and $F=300$kV/cm.
  }
\end{figure}

\begin{figure}[H]
  \includegraphics[width=250pt]{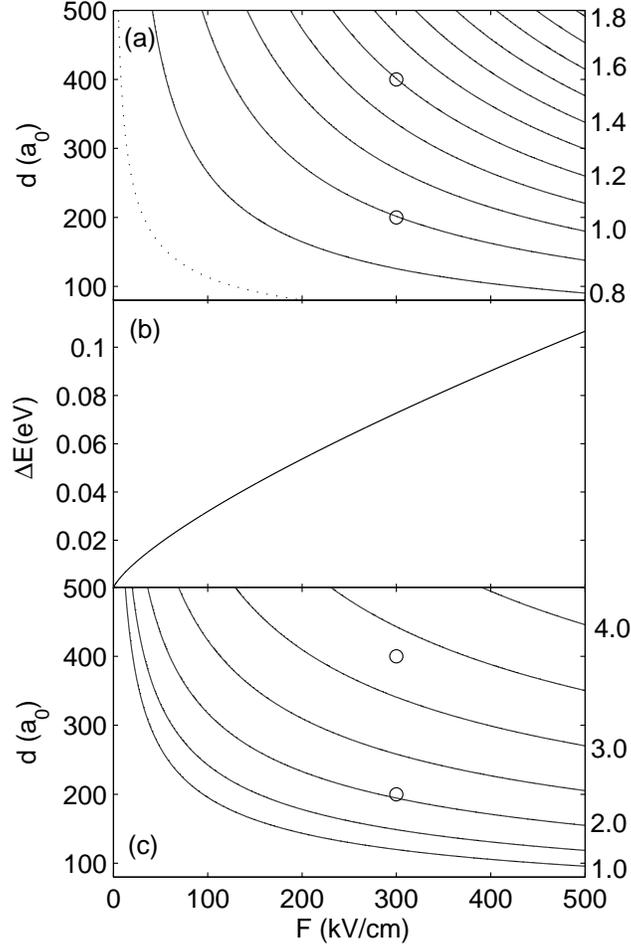}
  \caption{Estimating the barrier scattering region.
  (a) Barrier scattering window position in photon energy ($eV$) indicated on the
  right;
  (b) barrier scattering window width $\Delta E$ as a function of $F$ in Eq.(17);
  (c) number of oscillations within the barrier scattering window indicated on the right. The dotted line in (a) is the boundary of the region satisfying $F>1/(4d^2)$.  }
\end{figure}


\end{document}